\documentclass[prd,nofootinbib]{revtex4}
\usepackage{epsfig,amsmath}
\newcommand{\bra}[1]{\langle #1|}
\newcommand{\ket}[1]{|#1\rangle}
\newcommand{\braket}[2]{\langle #1|#2\rangle}

\def\d{{\rm d}}

\def\Pro{{\sf P}}

\newcommand{\tr}{{\rm tr}}

\newcommand{\e}{{\rm e}}

\newcommand{\omegav}{\boldsymbol{\omega}}
\newcommand{\betav}{\boldsymbol{\beta}}
\newcommand{\sigmav}{\boldsymbol{\sigma}}

\newcommand{\phiv}{\boldsymbol{\phi}}
\newcommand{\Piv}{\boldsymbol{\Pi}}

\newcommand{\p}{{\rm p}}
\newcommand{\x}{{\rm x}}

\newcommand{\poinc}{IO(1,3)$^\uparrow$}

\begin{document}

\title{The ideal relativistic spinning gas: polarization and spectra} 

\author{F. Becattini}\affiliation{Universit\`a di 
 Firenze and INFN Sezione di Firenze, Florence, Italy} 
\author{F. Piccinini}\affiliation{INFN Sezione di Pavia, Pavia, Italy}

\begin{abstract}
We study the physics of the ideal relativistic rotating gas at thermodynamical 
equilibrium and provide analytical expressions of the momentum spectra and 
polarization vector for the case of massive particles with spin 1/2 and 1. We 
show that the finite angular momentum ${\bf J}$ entails an anisotropy in momentum 
spectra, with particles emitted orthogonally to ${\bf J}$ having, on average, a 
larger momentum than along its direction. Unlike in the non-relativistic case, 
the proper polarization vector turns out not to be aligned with the total angular 
momentum with a non-trivial momentum dependence.
\end{abstract}

\maketitle
\section{Introduction}

The thermodynamics of rotating systems is a subject of interest mainly in astrophysics, 
where spinning objects are compact stars. Also in nuclear physics, there have
been noteworthy applications to the problem of multifragmentation \cite{moretto,gross}. 
In most those applications, either the non-relativistic approximation is used or the 
spin of particles is neglected, so that a full description of the relativistic 
case including spin degrees of freedom is still missing. However, recently, the 
microcanonical and grand-canonical partition function of an ideal relativistic quantum 
gas of particles with spin have been calculated~\cite{micro2} enforcing angular 
momentum conservation. Taking advantage of these results, in this paper we work out 
analytical expressions of the spectra and polarization of particles in a relativistic 
rotating gas with large (in $\hbar$ units) angular momentum. From a phenomenological
point of view, these calculations might be of interest for the physics of relativistic 
heavy ion collisions, where the formation of a system with a large intrinsic angular 
momentum has been envisaged \cite{satz}. We will confine ourselves to the case of 
Boltzmann statistics, leaving the quantum statistics case to future work.

The paper is organized as follows: in Sect.~\ref{rotating} we discuss the statistical 
mechanics of an ideal relativistic gas with fixed, large angular momentum and show 
the equivalence with a rigidly rotating system; in Sect.~\ref{moments} we analyze 
more in detail the relation between angular momentum and angular velocity in the 
limit of low rotational speed; in Sect.~\ref{spect} we calculate the inclusive 
momentum spectra of particles; finally, in Sect.~\ref{polariz} we obtain the 
expressions of the polarization vectors of massive particles with spin 1/2 and 1.

\subsection*{Notation}

In this paper we adopt the natural units, with $\hbar=c=K=1$. Space-time linear 
transformations (translations, rotations, boosts) and SL(2,C) transformations
are written in serif font, e.g. ${\sf R}$, ${\sf L}$. Operators in Hilbert space will
be denoted by an upper hat, e.g. $\widehat {\sf R}$. Unit vectors will be denoted 
with a smaller hat e.g. $\hat {\bf p}$. Even though the notation is unambiguous, 
we will make explicit mention of either possibility whenever confusion may arise. 

\section{Rotating relativistic gas at equilibrium}\label{rotating}

In general, by rotating thermodynamical system we mean a system with fixed angular
momentum in its rest frame. This is a good definition also in the relativistic case. 
It is well known that a classical system with non-vanishing intrinsic angular momentum 
can be at thermodynamical equilibrium only if the rotation is rigid \cite{landau}, that 
is if a constant vector $\omegav$ exists such that the local collective velocity 
${\bf v}$ is:
\begin{equation}\label{velocity}
  {\bf v} = \omegav \times  {\bf x}
\end{equation}
We will show that this conclusion holds for the relativistic gas, with the obvious 
inequality:
\begin{equation}\label{bound}
 \| \omegav \times {\bf x}\| < 1
\end{equation}
In this section, we will develop a proof based on an {\em ab initio} statistical 
mechanics calculation while a proof based on an extension of Landau's argument for
relativistic systems is given in Appendix B.

In ref.~\cite{micro2}, one of us derived the full expression of the microcanonical
partition function, as well as its grand-canonical limit for large volumes, of a 
multi-species ideal relativistic gas with fixed angular momentum. In relativistic 
quantum mechanics, fixing the angular momentum means projecting onto irreducible 
states of the orthochronous Poincar\'e group IO(1,3)$^\uparrow$ which, in fact, is 
a projection onto angular momentum states in the rest frame, where the total linear 
momentum vanishes. If, moreover, the intrinsic angular momentum is large 
(in $\hbar$ units), it can be treated as a classical vector ${\bf J}$ and the 
grand-canonical partition function reads \cite{micro2}:
\begin{eqnarray}\label{partfunc}
  Z_J &=&  \tr \{ \exp[(-\widehat H +\mu \widehat Q)/T] P_{\bf J} \Pro_V \} \nonumber \\
 &=& 
  \frac{2J + 1}{8\pi^2} \int_{|\phiv|<\pi} \!\!\!\! \d^3 \phi \; 
  \e^{i \phiv\cdot {\bf J}} \exp \left[ \sum_j \frac{\lambda_j}{(2\pi)^3}
  \int_V \d^3 \x \int \d^3 \p \; \e^{-\varepsilon_j/T} 
  \e^{-i \phiv \cdot({\bf x} \times {\bf p})} \, \tr D^{S_j}({\sf R}_{\hat{\phiv}}
  (\phi)) \right] 
\end{eqnarray}
where $T$ is the temperature and $V$ the volume; $\varepsilon_j$ is the energy, 
$\lambda_j$ the fugacity and $S_j$ the spin of the $j$th particle species; 
$D^{S_j}({\sf R}_{\hat{\phiv}}(\phi))$ is the matrix of the irreducible representation 
$S_j$ of SU(2) transformation ${\sf R}$ with axis $\hat{\phiv}$ and angle $\phi = \| \phiv \|$. 
The operator
$\widehat H$ is the hamiltonian, $\widehat Q$ is a generic conserved charge operator and 
$\mu$ the relevant chemical potential; the operator $\Pro_{\bf J}$ is the projector 
onto a fixed angular momentum while:
$$
 \Pro_V = \sum_{h_V} \ket{h_V}\bra{h_V}
$$
is the projector onto localized states $\ket{h_V}$, which form a complete set of quantum
states for the system in the finite region $V$. This projector is needed in 
eq.~(\ref{partfunc}) for the trace operation to be a properly defined one, i.e. 
involving a basis of the {\em full} Hilbert space \cite{micro2}.

The formula (\ref{partfunc}) applies to a system with large $V$ and $J$ and the integrand 
function turns out to be significant only in a region where $\phi = \| \phiv \| \ll 1$. 
Since both $J$ and $V$ are large, it is then possible to make a saddle-point expansion 
of the integral (\ref{partfunc}). To do this, we first have to continue the integration 
variables to the complex plane and, in view of the spherical symmetry of the domain, 
the obvious choice is to take the spherical coordinates of the vector $\phiv$,
namely its magnitude $\phi$ along with its polar and azimuthal angles. Then, we have 
to solve the complex vector equation:
\begin{equation}\label{saddlepoint}
 \nabla_{\phiv} \left[ i \phiv \cdot {\bf J} + \sum_j 
 \frac{\lambda_j}{(2\pi)^3} \int_V \d^3\x \int \d^3\p \, 
 \tr D^{S_j}({\sf R}_{\hat \phiv}(\phi)) \e^{-\varepsilon_j/T} 
 {\rm e}^{-i \phiv \cdot({\bf x} \times {\bf p})} \right] = 0 
\end{equation}
leading to:
\begin{eqnarray}\label{saddlepoint2}
{\bf J} &=& \sum_j \frac{\lambda_j}{(2\pi)^3} \int_V  {\rm d}^3\x
 \int {\rm d}^3\p \; {\rm tr} D^{S_j} ({\sf R}_{\hat \phiv}(\phi))  
 ({\bf x} \times {\bf p})\e^{-\varepsilon_j/T} \e^{-i \phiv \cdot({\bf x} 
 \times {\bf p})} \nonumber \\ 
&+& \sum_j \frac{\lambda_j}{(2\pi)^3} \int_V \d^3\x \int \d^3\p 
 \; \left[ i \nabla_{\phiv} \tr D^{S_j} ({\sf R}_{\hat \phiv}(\phi)) \right] 
 \e^{-\varepsilon_j/T} \e^{-i \phiv \cdot({\bf x} \times {\bf p})} \nonumber \\
&& \equiv {\bf L}(\phiv) + {\bf S}(\phiv)
\end{eqnarray}  
the reason for naming the two integral terms as ${\bf L}(\phiv)$ and ${\bf S}(\phiv)$
will become clear later on. We first observe that, if the equation (\ref{saddlepoint})
is solved by $\phi, \hat\phiv$, then also $-\phi^*, \hat \phiv^*$ is a solution.
This can be easily checked by taking into account that the trace depends only
on $\phi$ and can be written as a sum of exponentials $\exp[in\phi]$, being $n$
an integer. Therefore, we will look for one solution enforcing $\phi = -\phi^*$
and $\hat\phiv = \hat\phiv^*$, that is with a real unit vector $\hat\phiv$ and an
imaginary magnitude $\phi$. 

Assuming that one solution exists and defining the real vector $\omegav \equiv -iT\phiv$, 
we can write the grand-canonical partition function at the lowest order of the 
saddle-point expansion as:
\begin{equation}\label{partfunc2}
 Z_J \propto \exp[-\omegav \cdot {\bf J}/T] \exp \left[ \sum_j \frac{\lambda_j}
 {(2\pi)^3} \int_V {\rm d}^3\x \int {\rm d}^3\p \; {\rm tr} 
 D^{S_j}({\sf R}_{\hat{\omegav}}(i \omega/T)) {\rm e}^{-\varepsilon_j/T} 
 {\rm e}^{{\omegav}\cdot({\bf x} \times {\bf p})/T} \right]
\end{equation}
The function:
\begin{equation}\label{gpfomega}
 Z_\omega = \exp \left[ \sum_j \frac{\lambda_j}{(2\pi)^3} 
 \int_V {\rm d}^3\x \int {\rm d}^3\p \; {\rm tr} 
 D^{S_j}({\sf R}_{\hat\omegav}(i \omega/T)) {\rm e}^{-\varepsilon_j/T} 
 {\rm e}^{\omegav \cdot({\bf x} \times {\bf p})/T}\right]
\end{equation}
is the partition function of an ideal relativistic gas (in the Boltzmann limit) 
rotating with an angular velocity $\omegav$. This can be shown explicitely for the 
spinless case by dividing the gas into small cells with volume $\Delta^3 x$ and uniform 
velocity (\ref{velocity}) and calculating the relevant grand-canonical partition 
function in the formalism of relativistic thermodynamics \cite{touschek}:
\begin{equation}\label{cell}
 Z_{\rm cell} = \exp \left[ \sum_j \frac{\lambda_j}{(2\pi)^3} 
 \Delta^3\x \int {\rm d}^3\p \; {\rm e}^{-\beta \cdot p} \right] 
\end{equation}
where $\beta$ is the temperature four-vector:
\begin{equation}\label{beta}  
 \beta = \frac{1}{T_0 \sqrt{1-v^2}} (1, {\bf v}) = 
  \frac{1}{T_0 \sqrt{1-v^2}} (1, \omegav \times {\bf x})
\end{equation}
$T_0$ being the {\em local} temperature, i.e. the temperature measured by a 
thermometer moving along with the cell. Note that with this equation, we are tacitly
assuming that the partition function, which is a Lorentz-invariant quantity, in the
accelerated rotating cell is the same as that in an inertial frame with its instantaneous 
origin and velocity \footnote{Note that thermodynamic equilibrium is possible in a 
rotating cell because its acceleration is stationary}. This is consistent with 
the locality hypothesis \cite{mashhoon} and the transformation law of energy in 
a rotating frame \cite{mashhoon}:
\begin{equation}\label{noninert}
  E' = \gamma (E - \omegav \cdot {\bf L}) = \gamma (E - \omegav \cdot 
 ({\bf x} \times {\bf P}))
\end{equation}
being $\gamma = 1/\sqrt{1- \|\omegav \times {\bf x}\|^2}$. Therefore, with the 
identification:
\begin{equation}\label{temp}
  T = T_0 \sqrt {1-v^2} = T_0 \sqrt {1-\| \omegav \times {\bf x} \|^2} 
\end{equation}
the partition function of the cell becomes:
\begin{equation}\label{cell2}
 Z_{\rm cell} = \exp \left[ \sum_j \frac{\lambda_j}{(2\pi)^3} 
 \Delta^3\x \int {\rm d}^3\p \; {\rm e}^{-\varepsilon_j/T} 
 {\rm e}^{{\bf p} \cdot(\omegav \times {\bf x})/T} \right] 
\end{equation}
The total partition function of the supposedly independent cells can now be obtained 
by multiplying expressions (\ref{cell2}) for all cells and going to the limit of 
infinitesimal cells, implying:
\begin{equation}\label{celltotal}
 Z_\omega = \exp \left[ \sum_j \frac{\lambda_j}{(2\pi)^3} 
 \int_V {\rm d}^3\x \int {\rm d}^3\p \; {\rm e}^{-\varepsilon_j/T} 
 {\rm e}^{\omegav \cdot({\bf x} \times {\bf p})/T} \right] 
\end{equation}
which is the (\ref{gpfomega}) for spinless particles.

Therefore, unlike in the non-relativistic case, the proper temperature $T_0$ in a 
spinning relativistic system at equilibrium depends on the distance from the 
rotation axis, an effect pointed out by Israel \cite{israel}. However, the thermal 
equilibrium is characterized by a uniform temperature $T$ as measured by the observer;
another proof of this statement based on simple arguments can be found in Appendix B.

The partition function $Z_\omega$ defines an ensemble, that we can define as {\em 
rotational grand-canonical} where the angular velocity is fixed and the intrinsic
angular momentum ${\bf J}$ can fluctuate. On the other hand, the grand-canonical 
partition function with fixed intrinsic angular momentum ${\bf J}$ defines an ensemble 
that can be called {\em micro-rotational grand-canonical}. The above definitions 
are inspired of the more familiar canonical-microcanonical duality, according to 
whether $T$ or $E$ are fixed. 

Pursuing the analogy, it is expected that these two ensembles become equivalent in 
the thermodynamic limit as far as the calculation of first-order moments of statistical
distributions is concerned. In fact, neglecting the constant small factors multiplying  
$Z_J$, it can be seen from (\ref{partfunc2}) that $\log Z_\omega$ is essentially 
the Legendre transform of $\log Z_J$ with respect to the total angular momentum ${\bf J}$:
\begin{equation}\label{legendre}
 \log Z_\omega = \log Z_J + \frac{\omegav \cdot {\bf J}}{T} = \log Z_J + 
  {\bf J} \cdot \frac{\partial}{\partial {\bf J}} \log Z_J
\end{equation}
The saddle-point equation (\ref{saddlepoint2}) for $\phiv = i \omegav/T $ is simply 
the inverse Legendre transformation of (\ref{legendre}):
\begin{eqnarray}\label{saddlepoint3} 
 {\bf J} &=&  \sum_j \frac{\lambda_j}{(2\pi)^3} \int_V {\rm d}^3\x 
 \int {\rm d}^3\p \; {\rm tr} D^{S_j}({\sf R}_{\hat\omegav}(i \omega/T))
 ({\bf x} \times {\bf p}) {\rm e}^{-\varepsilon_j/T} 
 {\rm e}^{\omegav \cdot({\bf x} \times {\bf p})/T} \nonumber \\
 &+& \sum_j \frac{\lambda_j}{(2\pi)^3} \int_V {\rm d}^3\x \int {\rm d}^3\p\; 
  \left[ \frac{\partial}{\partial \omega/T} {\rm tr} 
  D^{S_j}({\sf R}_{\hat\omegav}(i \omega/T)) \right] \hat\omegav
  {\rm e}^{-\varepsilon_j/T} {\rm e}^{\omegav \cdot({\bf x} \times {\bf p})/T} 
\nonumber \\
 &=& {\bf L}(\omegav/T) + {\bf S}(\omegav/T) = \frac{\partial}{\partial {\omegav/T}} 
 \log Z_\omega 
\end{eqnarray}
Therefore, the saddle point equation (\ref{saddlepoint3}) relates angular velocity 
$\omegav$ and angular momentum ${\bf J}$ and applies to relativistic as well as 
non-relativistic gases. It expresses the conservation of angular momentum, in the 
sense that the total angular momentum in the rotational grand-canonical ensemble 
on the right hand side, where the two contributions of orbital and spin can be 
clearly identified, equals the initial conserved total angular momentum ${\bf J}$ 
on the left hand side. It is worth pointing out that for macroscopic systems,
the contribution of the orbital angular momentum is predominant. This can be seen
from eq.~(\ref{saddlepoint3}) noting that the integral defining the total orbital
angular momentum is scaled by a factor $\| {\bf x} \times {\bf p} \|$ with 
respect to the spin angular momentum. Since this is generally much larger than 1,
for macroscopic distances and not too low temperatures, we usually have $\| {\bf L}\|
\gg \| {\bf S} \|$ and the spin angular momentum can be neglected. Yet, for not
too large systems, the spin angular momentum contribution can be sizeable.
There is also a remarkable difference between them concerning their dependence
on the size of the sytem: the spin angular momentum is properly {\em extensive},
i.e. it increase linearly with the volume, while the orbital angular momentum is not
because of the factor $\| {\bf x} \times {\bf p} \|$. The lack of extensivity of
rotating systems makes the thermodynamic limit essentially irrelevant and 
dictates a finite-size treatment. From a completely equivalent point of view
of the comoving frame, we can state that the rotating system is non-extensive 
because of the presence of a long-range centrifugal potential.

Finally, it can be shown that in the rotational grand-canonical ensemble each
microstate with total energy $E$, total charge $Q$ and total angular momentum ${\bf J}$ 
has a probability (in the frame where $\langle{\bf P}\rangle=0$):
\begin{equation}\label{proba}
    p(T,\mu,\omegav) \propto \exp[-E/T + \mu Q/T + \omegav \cdot {\bf J}/T]
\end{equation}
and, consequently, rotational grand-canonical partition function can be written as:
\begin{equation}\label{trace}
 Z_\omega = \tr \{ 
 \exp [(-\widehat H + \mu \widehat Q + \omegav \cdot \widehat{\bf J})/T] \Pro_V \}
\end{equation}
where $\widehat {\bf J}$ is the total angular momentum operator.
This statement holds provided that $\omega/T \ll 1$ so that terms of the order
$(\omega/T)^2$ and higher can be neglected; only in this case is the formula (\ref{trace}) 
equivalent to (\ref{gpfomega}). Indeed, working out the trace (\ref{trace}) for an 
ideal gas involves the calculation of matrix elements such as:
\begin{equation}\label{complexrot}
 \sum_{\sigma} \bra{p,\sigma}\exp[\omegav \cdot \widehat{\bf J}/T] \Pro_V \ket{p,\sigma} =
 \sum_{\sigma} \bra{p,\sigma} \hat{\sf R}_{\hat\omegav}(i\omega/T) \Pro_V \ket{p,\sigma}
\end{equation}
where $\ket{p,\sigma}$ are single particle states with four-momentum $p$ and spin
projection $\sigma$ (see Sect.~5 for extended discussion) and the rotation of an imaginary 
angle $i\omega/T$ has been introduced. The above single-particle trace has been 
calculated in ref.~\cite{micro2}:
\begin{equation}\label{matrix}
 \sum_\sigma \bra{p,\sigma} \hat{\sf R}_{\hat\omegav}(i\omega/T] \Pro_V \ket{p,\sigma}
= \frac{1}{(2\pi)^3} \int_V \d^3 \x \; \e^{i {\bf x}\cdot({\bf p}-
 {\sf R}_{\hat \omegav}(i\omega/T)^{-1}{\bf p})} \tr D^S ({\sf R}_{\hat \omegav}(i\omega/T))
 \bra{0} \Pro_V \ket{0} 
\end{equation}
The vacuum expectation value of $\Pro_V$ is immaterial and also becomes 1 in the 
large volume limit. Now, since:
\begin{equation}\label{rotazvett}
 {\sf R}_{\hat \omegav}(\psi){\bf v}={\bf v} \cos \psi  + (\hat{{\bf n}} 
 \times {\bf v}) \sin \psi +(1- \cos \psi ){\bf v}\cdot\hat{{\bf n}}\,\hat{{\bf n}}
\end{equation}
we have:
$$
  {\bf x}\cdot({\bf p}-{\sf R}_{\hat \omegav}(i\omega/T)^{-1}{\bf p})]
 \simeq i \frac{\omega}{T}{\bf x} \cdot (\omegav \times {\bf p})
$$
only if $\omega/T$ is sufficiently smaller than 1, so that cosine terms in 
(\ref{rotazvett}) can be approximated with 1 and sine terms with their argument.
In this case, eq.~(\ref{matrix}) turns into, for large $V$:
\begin{equation}
 \sum_\sigma \bra{p,\sigma} \hat{\sf R}_{\hat\omegav}(i\omega/T) \Pro_V \ket{p,\sigma}
 \simeq \frac{1}{(2\pi)^3} \int_V \d^3 \x \; \e^{\omegav \cdot ({\bf x}\times{\bf p})/T}
 \tr D^S ({\sf R}_{\hat \omegav}(i\omega/T))
\end{equation}
It is then straightforward to show that the final expression of (\ref{trace}) 
is just (\ref{gpfomega}).

\section{Moments of inertia}\label{moments}

In this section, we will study more in detail the saddle-point equation 
(\ref{saddlepoint2}) or its equivalent form (\ref{saddlepoint3}). This is a vector 
equation in $\omegav$, whose solution depends, besides $T$ and ${\bf J}$, on the 
shape and size of the system. 

If the total angular momentum ${\bf J}$ is a simmetry axis of $V$, then ${\bf J}$ and 
$\omegav$ must be parallel (see Fig.~1). This can be proved by applying an arbitrary rotation 
of an angle $\psi$ around ${\bf J}$, let ${\sf R}_{\hat{\bf J}}(\psi)$, to both sides 
of eq.~ (\ref{saddlepoint3}). Since ${\sf R}({\bf x} \times {\bf p}) = 
{\sf R}{\bf x} \times {\sf R}{\bf p}$, defining new integration variables 
${\bf x}'={\sf R}{\bf x}$ and ${\bf p}'={\sf R}{\bf p}$, taking into account
that $\omegav \cdot{\sf R}^{-1}({\bf x}' \times {\bf p}') = {\sf R}\omegav \cdot
({\bf x}' \times {\bf p}')$ and being both momentum and spacial domains invariant
by rotation around ${\bf J}$, we can conclude that, if $\omegav$ is a solution of
the eq.~(\ref{saddlepoint}), ${\sf R}_{\hat{\bf J}}(\psi)\omegav$ is also a solution
{\em for any} $\psi$. Therefore, ${\sf R}_{\hat{\bf J}}(\psi)\omegav$ should coincide 
with $\omegav$ for the equation to be well behaved. If this is the case, $\omegav$ is
parallel to ${\bf J}$ and the orbital term ${\bf L}(\omegav/T)$ should be also
parallel to $\omegav$. In this case, the eq.~(\ref{saddlepoint3}) essentially reduces
to a scalar one:
\begin{eqnarray}\label{saddlepoint4} 
 && J  =  \sum_j \frac{\lambda_j}{(2\pi)^3} \int_V {\rm d}^3\x 
 \int {\rm d}^3\p \; {\rm tr} D^{S_j}({\sf R}_{\hat\omegav}(i \omega/T))
 \hat{\bf J} \cdot({\bf x} \times {\bf p}) {\rm e}^{-\varepsilon_j/T} 
 {\rm e}^{\omega \hat{\bf J} \cdot({\bf x} \times {\bf p})/T} \nonumber \\
 && + \sum_j \frac{\lambda_j}{(2\pi)^3} \int_V {\rm d}^3\x \int {\rm d}^3\p\; 
  \left[ \frac{\partial}{\partial \omega/T} {\rm tr} 
  D^{S_j}({\sf R}_{\hat\omegav}(i \omega/T)) \right]
  {\rm e}^{-\varepsilon_j/T} {\rm e}^{\omega \hat{\bf J} \cdot({\bf x} \times {\bf p})/T}
\end{eqnarray}
which can be solved numerically.

Let us now introduce a length $R$ related to the size of the system; it can be, 
e.g. the maximal distance of a point of the set $V$ from the rotation axis. 
An analytic solution of the general equation (\ref{saddlepoint3}) can be obtained 
for small values of $\omega R \p/T$. In fact, under this circumstance, the 
exponential can be approximated as:
\begin{equation}\label{approx}
 \exp[\omegav \cdot({\bf x} \times {\bf p})/T] 
 \simeq 1 + \omegav \cdot({\bf x} \times {\bf p})/T
\end{equation}
This condition leads to different requirements for a non-relativistic and an 
ultrarelativistic gas (being $p = {\cal O}(\sqrt{mT})$ with $m \gg T$ and 
$p={\cal O}(T)$ respectively), yet both imply that $\omega R \ll 1$. If approximation 
(\ref{approx}) applies, the two terms ${\bf L}$ and ${\bf S}$ reduce to, at the 
first order in $\omega$:
\begin{eqnarray}
&& {\bf L} \simeq \sum_j \frac{\lambda_j}{(2\pi)^3} \int_V {\rm d}^3\x 
 \int {\rm d}^3\p \; (2S_j+1)({\bf x} \times {\bf p}) \frac{\omegav}{T} 
 \cdot({\bf x} \times {\bf p}) {\rm e}^{-\varepsilon_j/T} \nonumber \\
&& {\bf S} \simeq \sum_j \frac{\lambda_j}{(2\pi)^3} \int_V {\rm d}^3\x 
  \int {\rm d}^3\p\; \frac{S_j(S_j+1)(2S_j+1)}{3}
 {\rm e}^{-\varepsilon_j/T} \frac{\omegav}{T}
\end{eqnarray}
By expanding the scalar and vector products in the integral of the orbital angular
momentum and carrying out simple algebraic calculations, one obtains:
\begin{equation}\label{lomega}
 L_i = \sum_{k=1}^3 \sum_j \langle n_j \rangle \frac{\langle p_j^2 \rangle}{3T} 
 \int_V {\rm d}^3\x \; (r^2 \delta_{ik} - \x_i \x_k ) \; \omega_k
\end{equation}
where $r^2= \sum_{i=1}^3 \x^2_i$ and:
\begin{equation}
\langle n_j \rangle = \frac{\lambda_j (2S_j + 1)}{(2\pi)^3} \int \d^3 \p \;
 {\rm e}^{-\varepsilon_j/T}
\end{equation}
is the mean density of the species $j$ in the Boltzmann approximation and without
angular momentum constraint, while:
\begin{equation}
\langle p_j^2 \rangle = \frac{1}{\langle n_j \rangle}
 \frac{\lambda_j (2S_j + 1)}{(2\pi)^3} \int \d^3 \p \;
 \p^2 {\rm e}^{-\varepsilon_j/T}
\end{equation}
is the mean squared momentum of the particle $j$ under the same approximations.

The relation between ${\bf L}$ and $\omegav$ (\ref{lomega}) resembles the 
classical linear relation between angular velocity and angular momentum defining 
the inertia tensor ${\sf I}$. Indeed, in the non-relativistic limit, this is 
precisely what one gets from (\ref{lomega}) because $\langle p_j^2 \rangle = 3 m_j T$ 
and, consequently:
\begin{equation}\label{lomega2}
  L_i = \sum_{k=1}^3 \sum_j m_j \langle n_j \rangle 
 \int_V {\rm d}^3\x \; (r^2 \delta_{ik} - {\rm x_i}{\rm x_k}) \; \omega_k 
 = \sum_{k=1}^3 I_{ik} \omega_k
\end{equation}
Hence the expression in (\ref{lomega}):
\begin{equation}
  I_{ik} = \sum_j \langle n_j \rangle \frac{\langle p_j^2 \rangle}{3T} 
 \int_V {\rm d}^3\x \; (r^2 \delta_{ik} - \x_i \x_k )
\end{equation}
turns out to be the right generalization of the inertia tensor for a relativistic
gas. Like in the classical case, if the system (which is homogeneous for small $\omega$)
is symmetric around its rotation axis, then ${\bf L}$ is parallel to $\omegav$ and 
the proportionality constant is just the moment of inertia with respect to that axis. 

Unlike in the classical case, for non-macroscopic systems there is also a potentially 
non-negligible spin contribution. In this case, the full relation between ${\bf J}$
and $\omegav$ reads:
\begin{equation}\label{jomega}
 J_i = \sum_{k=1}^3 \sum_j \langle n_j \rangle \left[ 
 \frac{\langle p_j^2 \rangle}{3T} \int_V {\rm d}^3\x \; 
 (r^2 \delta_{ik} - \x_i \x_k ) + V \frac{S_j(S_j+1)}{3T} \delta_{ik} 
 \right] \; \omega_k
\end{equation}
%
\begin{center}
\begin{figure}[ht]
\epsfxsize=4.5in
\epsffile{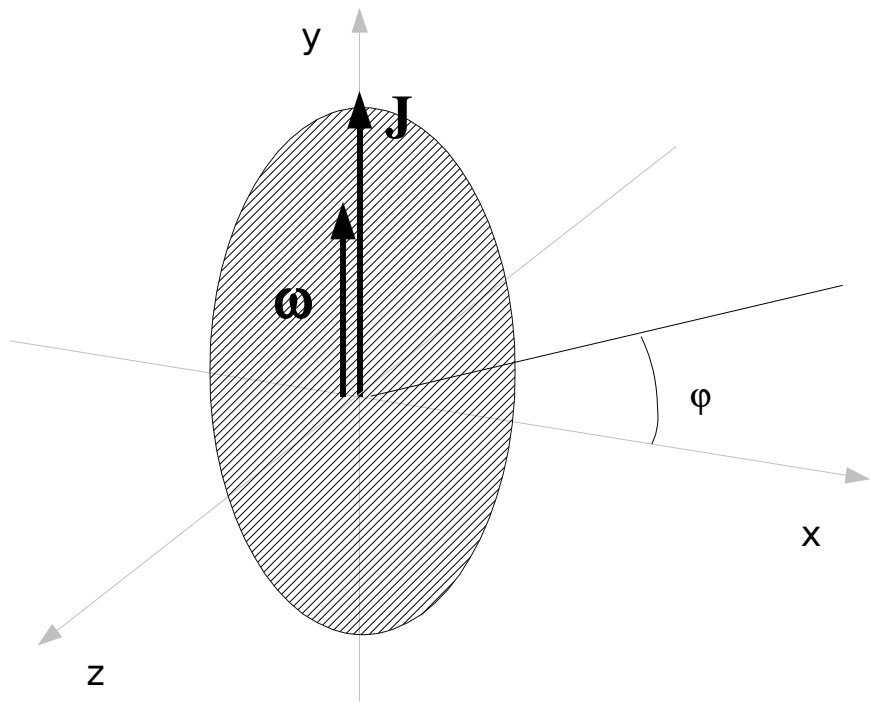} 
\caption{For an  axially symmetric system around the angular momentum direction,
$\omegav$ is parallel to ${\bf J}$. Also shown the coordinates chosen to describe
spectra in Sect.~\ref{spect}.
\label{system}}
\end{figure}
\end{center} 

\section{Spectra and anisotropies}\label{spect}

The impact of a finite angular momentum on particle spectra in a relativistic
gas was considered by Hagedorn \cite{hagedorn} who pointed out that a large value of the
intrinsic angular momentum involves a peculiar anisotropy. Here, we will provide
the most general expressions of the momentum spectrum of particles in an equilibrated 
rotating system with fixed (classical) angular momentum ${\bf J}$. This can be 
obtained directly, for the $j$th particle species, from the partition function 
(\ref{partfunc}):
\begin{equation}
 \left\langle \frac{\d n_j}{\d^3 \p}\right\rangle = \frac{1}{Z_J} \tr 
 \left\{ \frac{\d \widehat n_j}{\d^3 \p} 
 \exp[(-\widehat H +\mu \widehat Q)/T] P_J \Pro_V \right\} \nonumber \\
\end{equation}
where $\d \widehat n_j/\d^3 \p$ is the momentum spectrum {\em operator}. 
The above expression can be rewritten as the functional derivative with
respect to $\alpha({\bf p})$ of:
\begin{equation}\label{functional}
  Z_J[\alpha({\bf p})] = 
  \tr \left\{ \exp \left[(-\widehat H +\mu \widehat Q)/T + \int \d^3 \p \; 
  \alpha({\bf p}) \frac{\d \widehat n_j}{\d^3 \p} \right] P_{\bf J} \Pro_V \right\} 
\end{equation}
in $\alpha({\bf p}) = 0$. Since, for an ideal gas:
$$
 \widehat H = \sum_j \int \d^3 \p \; \varepsilon_j({\bf p}) \frac{\d \widehat n_j}{\d^3 \p}
$$
the functional $Z_J[\alpha({\bf p})]$ can be simply obtained from eq.~(\ref{partfunc})
replacing $\varepsilon_j/T$ with $\varepsilon_j/T + \alpha({\bf p})$. The functional
derivative then turns out to be:
\begin{eqnarray}\label{genspect}
  \left\langle \frac{\d n_j}{\d^3 \p}\right\rangle 
  &=& \frac{\delta}{\delta \alpha({\bf p})} 
  \log Z_J[\alpha({\bf p})]\Big|_{\alpha({\bf p})=0} = 
  \frac{1}{Z_J} \frac{2J + 1}{8\pi^2} 
  \frac{\lambda_j}{(2\pi)^3} \int_V \d^3 \x \; \e^{-\varepsilon_j/T} 
  \int_{|\phiv|<\pi} \!\!\!\! \d^3 \phi \; 
  \e^{i \phiv\cdot ({\bf J}-\x \times \p)} \tr D^{S_j}({\sf R}_{\hat{\phiv}}(\phi))
 \nonumber \\
  &\times& 
  \exp \left[ \sum_k \frac{\lambda_k}{(2\pi)^3}
  \int_V \d^3 \x \int \d^3 \p \; \e^{-\varepsilon_k/T} 
  \e^{-i \phiv \cdot({\bf x} \times {\bf p})} \, 
  \tr D^{S_k}({\sf R}_{\hat{\phiv}}(\phi)) \right] 
\end{eqnarray}
Similarly to eq.~(\ref{partfunc}), it is possible to make a saddle-point expansion
of the above integral. If the orbital angular momentum of the particle is much lower
than the total angular momentum, i.e. $\| {\bf x} \times {\bf p} \| \ll J$, the former can be 
neglected in the saddle-point equation which is then essentially the same as that 
for the partition function, eq.~(\ref{saddlepoint}). This condition amounts to take
$\exp[-i \phiv \cdot ({\bf x} \times {\bf p})]$ as a constant factor and it is usually met 
except for momenta larger than $J/R$, $R$ being the size of the system. 
In this case, the formula (\ref{genspect}) becomes, at the leading order:
\begin{equation}\label{spectrum}
  \left\langle \frac{\d n_j}{\d^3 \p}\right\rangle 
  = \frac{\lambda_j}{(2\pi)^3}\tr D^{S_j}({\sf R}_{\hat{\omegav}}(i\omega/T))
    \int_V \d^3 \x \; \e^{-\varepsilon_j/T} \e^{\omegav \cdot ({\bf x} \times {\bf p})/T} 
\end{equation}
which is, of course, the spectrum in the rotational grand-canonical ensemble in
the Boltzmann limit. 

It is of some interest to derive the shape of the spectra in cylindrical 
momentum coordinates $p_T = \sqrt{p_x^2 + p_y^2}$, azimuthal angle $\varphi$
and rapidity $y = 1/2 \log [(p_z + \varepsilon)/(\varepsilon-p_z)]$ when the
angular momentum of the system is directed along the $y$ axis, i.e. ${\bf J} = J 
\hat{\bf j}$ and the system is rotationally symmetric around ${\bf J}$. Because
of this symmetry one has $\omegav = \omega \hat{\bf j}$ and the spectrum 
(\ref{spectrum}) reads:
\begin{equation}\label{spectrum1}
  \left\langle \frac{\d n_j}{\d^3 \p}\right\rangle 
  = \frac{\lambda_j}{(2\pi)^3}\tr D^{S_j}({\sf R}_{\hat{\omegav}}(i\omega/T))
    \int_V \d^3 \x \; \e^{-\varepsilon_j/T} \e^{\omegav \cdot (\x \times \p)/T} 
\end{equation}
In the new coordinates $p_T,\varphi,y$ the spectrum reads:
\begin{equation}\label{spectrum2}
  \left\langle \frac{\d n_j}{p_T \d p_T \d \varphi \d y} \right\rangle 
  = \frac{\lambda_j\tr D^{S_j}({\sf R}_{\hat{\omegav}}(i\omega/T))}{(2\pi)^3}
    \int_V \d^3 \x \; m_T \cosh y \exp [-m_T \cosh y/T + 
     {\bf p}_T \cdot (\omegav \times {\bf x})_\perp/T + m_T 
     (\omegav \times {\bf x})_\parallel \sinh y/T ] 
\end{equation}
where $m_T=\sqrt{p_T^2+m_j^2}$, $_\parallel$ labels the longitudinal projection
along the $z$ axis and $_\perp$ the transverse projection onto the $xy$ plane.

With successive integrations of eq.~(\ref{spectrum2}) one obtains:
\begin{equation}\label{spectrumphi}
  \left\langle \frac{\d n_j}{p_T \d p_T \d \varphi} \right\rangle 
  = \frac{\lambda_j\tr D^{S_j}({\sf R}_{\hat{\omegav}}(i\omega/T))}{4\pi^3}
    \int_V \d^3 \x \; \frac{1}{\sqrt{1-(\omegav \times {\bf x})_\parallel^2}}
      m_T {\rm K}_1 \left( m_T \sqrt{1-(\omegav \times {\bf x})_\parallel^2}/T \right)
     \exp [{\bf p}_T \cdot (\omegav \times {\bf x})_\perp/T] 
\end{equation}
and:
\begin{equation}\label{spectrumpt}
  \left\langle \frac{\d n_j}{p_T \d p_T} \right\rangle 
  = \frac{\lambda_j\tr D^{S_j}({\sf R}_{\hat{\omegav}}(i\omega/T))}{2\pi^2}
    \int_V \d^3 \x \; \frac{1}{\sqrt{1-(\omegav \times {\bf x})_\parallel^2}}
      m_T {\rm K}_1 \left( m_T \sqrt{1-(\omegav \times {\bf x})_\parallel^2}/T \right)
      {\rm I}_0 (p_T \| (\omegav \times {\bf x})_\perp \|/T)  
\end{equation}
where ${\rm K}_1$ and ${\rm I}_0$ are McDonald and modified Bessel functions 
respectively. It is also possible to calculate the mean multiplicity by integrating 
(\ref{spectrumpt}), which yields:
\begin{equation}\label{mult}
  \langle n_j\rangle 
  = \frac{\lambda_j}{2\pi^2}\tr D^{S_j}({\sf R}_{\hat{\omegav}}(i\omega/T))
    \int_V \d^3 \x \; \frac{m^2 T}{1-\|\omegav \times {\bf x}\|^2}
     {\rm K}_2 \left( m\sqrt{1-\|\omegav \times {\bf x}\|^2}/T \right)  
\end{equation}
The formulae (\ref{spectrum2})-(\ref{mult}) yield as limiting cases for $\omegav=0$
the well known ones for the relativistic ideal Boltzmann gas. 

It is worth pointing out that the spectrum (\ref{spectrum2}) features a non-trivial 
dependence on the momentum azimuthal angle $\varphi$ strongly sensitive to the ratio 
$\omega/T$. Introducing cylindrical coordinates $r,\Phi,z$ for the vector ${\bf x}$ 
and taking into account that $\omegav = \omega \hat{\bf j}$, eq.~(\ref{spectrumphi}) 
can be written as:
\begin{equation}\label{spectrumphicyl}
  \left\langle \frac{\d n_j}{p_T \d p_T \d \varphi} \right\rangle 
  = \frac{\lambda_j\tr D^{S_j}({\sf R}_{\hat{\omegav}}(i\omega/T))}{4\pi^3}
    \int_V \d^3 \x \; \frac{1}{\sqrt{1-(\omega r \cos \Phi)^2}}
      m_T {\rm K}_1 \left( m_T \sqrt{1-(\omega r \cos \Phi)^2}/T \right)
     \exp [ p_T \omega z \cos \varphi/T] 
\end{equation}
which shows that particles emitted along the momentum axis (with $\varphi=\pi/2$)
have, on average, a lower momentum than those emitted orthogonally to it.
Classically, this can be understood as an effect of centrifugal force.

The (\ref{spectrumphicyl}) is a periodic function in the momentum azimuthal angle
$\varphi$, thus it can be Fourier expanded. Borrowing a notation commonly used in
relativistic heavy ion physics, we will denote by $2v_k$ the coefficients of
this expansion, i.e. setting:
\begin{equation}\label{fseries}
  \left\langle \frac{\d n_j}{p_T \d p_T \d \varphi} \right\rangle 
  = \frac{1}{2\pi} \left\langle \frac{\d n_j}{p_T \d p_T} \right\rangle \sum_{k=0}^\infty
    2v_{2k} \cos 2 k \varphi 
\end{equation}
with $v_0 \equiv 1$ we obtain:
\begin{equation}\label{v2k}
 v_{2k} = \frac{\int_0^{2\pi} \left\langle \frac{\d n_j}{p_T \d p_T \d \varphi} 
  \right\rangle \cos 2 k \varphi} {\int_0^{2\pi} \left\langle \frac{\d n_j}{p_T \d p_T 
  \d \varphi} \right\rangle} 
\end{equation}
and, by using (\ref{spectrumphicyl})
\begin{equation}\label{v2kfinal}
 v_{2k} = \frac{\int_V \d^3 {\rm x} \; \frac{1}{\sqrt{1-(\omega r \cos \Phi)^2}}
      m_T {\rm K}_1 \left( m_T \sqrt{1-(\omega r \cos \Phi)^2}/T \right) 
      I_{2k}(p_T z \omega/T)}{\int_V \d^3 {\rm x} \; 
      \frac{1}{\sqrt{1-(\omega r \cos \Phi)^2}}
      m_T {\rm K}_1 \left( m_T \sqrt{1-(\omega r \cos \Phi)^2}/T \right) 
      I_{0}(p_T z \omega/T)}
\end{equation}
which shows that the Fourier coefficients are bound between 0 and 1. For low
$p_T$, the ratio between modified Bessel functions is such that ${\rm I}_{2k}/
{\rm I}_0 \propto p_T^{2k}$, while at large $p_T$ values, the ratio of modified
Bessel functions tends to 1 and, as a consequence, all $v_{2k} \to 1$.

\section{Polarization}\label{polariz}

If a gas has a net electric charge, an imbalance in the mean multiplicities of positive 
and negative particles is implied. Similarly, if a gas at thermodynamical 
equilibrium has a non-vanishing angular momentum (hence it is rigidly rotating, as 
we have seen in Sect.~\ref{rotating}) particles should have a net polarization along the 
direction of angular momentum. From the viewpoint of the comoving observer in
the rotating frame, the hamiltonian in the rotating frame has a spin-rotation 
coupling term \cite{mashhoon,quantrot} \footnote{This phenomenon is well known for radio
waves \cite{mashhoon}}. For the non-relativistic case, the calculation is 
straightforward as the polarization vector is the same in the observer and in the 
comoving frame, hence it can be carried out in the latter where the single-particle 
hamiltonian reads \cite{quantrot}:
\begin{equation}
 \widehat h_{\rm rot} = \widehat h_{\rm obs} - \omegav \cdot \widehat{\bf j}  
\end{equation}
being $\widehat{\bf j}$ the {\em total} (orbital + spin) angular momentum operator
of the particle. In fact, the problem we want to solve here is to calculate 
the polarization vector for a {\em relativistic} gas.

This problem is more difficult than for charge, energy or momentum because 
angular momentum is not a generator of an abelian group and this adds some 
complication. Particularly, from the point of view of statistical mechanics, one 
has to deal, as we will shortly see, with a spin density matrix with non-vanishing 
off-diagonal elements, unlike for momenta. From eq.~(\ref{trace}), the density 
operator in the rotational grand-canonical ensemble reads:
\begin{equation}\label{rota}
 \widehat\rho_\omega = 
 \frac{1}{Z_\omega} \exp [(-\widehat H + \mu \widehat Q + \omegav \cdot \widehat {\bf J})/T]
 \Pro_V
\end{equation}
while in the micro-rotational grand-canonical ensemble, from eq.~(\ref{partfunc}):
\begin{equation}\label{microrota}
 \widehat\rho_J = \frac{1}{Z_J} \exp [(-\widehat H + \mu \widehat Q)/T] \Pro_{\bf J} 
 \Pro_V
\end{equation}
$\Pro_{\bf J}$ being the quantum projector onto states with definite angular 
momentum. As argued in Sect.~\ref{rotating}, the two ensembles are equivalent for 
large $J$ and $V$ provided that $\omega/T$ is sufficiently smaller than 1. Henceforth, 
we will confine our attention to this case and we will work in the rotational 
grand-canonical ensemble described by (\ref{rota}). 

In the Boltzmann limit of the ideal relativistic quantum gas, all particles can 
be handled as independent distinguishable objects. Thus, the density operator 
(\ref{rota}) factorize and we can calculate the polarization by considering the 
single-particle density operator:
\begin{equation}\label{rotasingle}
 \widehat\rho_\omega = 
 \frac{1}{z_\omega} \exp [(-\widehat h + \mu \widehat q + \omegav \cdot \widehat{\bf j})/T]
 \Pro_V
\end{equation}
where $z_\omega$ is the single particle partition function:
\begin{equation}
  z_\omega = \frac{\lambda}{(2\pi)^3} \int_V \d^3 \x \int \d^3 \p \; 
  \e^{-\varepsilon/T} \e^{\omegav \cdot({\bf x} \times {\bf p})/T} \, 
  \tr D^{S}({\sf R}_{\hat{\omegav}} (i\omega/T)) 
\end{equation}
ensuring the correct normalization $\tr \rho_\omega = 1$ (this in fact requires
$\omega/T \ll 1$, see discussion at the end of Sect.~\ref{rotating}).

The density operator restricted to the spin degrees of freedom is a function of
momentum and reads:
\begin{equation}\label{rhop}
  \widehat\rho_\omega (p) = 
  \frac{1}{\frac{\lambda}{(2\pi)^3} \int_V \d^3 \x \; 
  \e^{-\varepsilon/T} \e^{\omegav \cdot({\bf x} \times {\bf p})/T} \, 
   \tr D^{S}({\sf R}_{\hat{\omegav}} (i\omega/T))} 
   \exp [(-\widehat h + \mu \widehat q + \omegav \cdot \widehat{\bf j})/T] \Pro_V
\end{equation}
The polarization is, by definition, the trace of the density matrix $\rho(p)$ multiplied
by a suitable spin operator. In non-relativistic quantum mechanics, this is obviously
the spin vector operator $\widehat{\bf S}$ and the polarization can be obtained straightforwardly: 
\begin{equation}\label{nonrela}
  {\bf \Pi} = \tr [{\widehat{\bf S} \widehat \rho_\omega(p)}] = 
  \frac{\sum_{n=-S}^S n \e^{n\omega/T}}
  {\sum_{n=-S}^S \e^{n\omega/T}} \hat \omegav
\end{equation}
which turns out to be momentum-independent.
In relativistic quantum mechanics the proper generalization of the spin angular
momentum is the Pauli-Lubanski vector multiplied by $1/m$:
\begin{equation}\label{luba}
\widehat W_\mu = -(1/2) \sum_{\nu \rho \sigma} 
 \epsilon_{\mu \nu \rho \sigma} \widehat J^{\nu \rho} 
 \widehat P^\sigma 
\end{equation} 
which in fact reduces to the spin operator in the particle rest frame; $\widehat J^{\nu \rho}$ 
are the generators of the Lorentz group. The polarization is thus a four-vector $\Pi$:
\begin{equation}\label{pola} 
  \Pi(p) = \frac{1}{m} \tr_p ( \widehat W \widehat \rho(p))
\end{equation}
where the trace is to be calculated by summing only over spin degrees of freedom
keeping the four-momentum $p$ fixed. This four-vector $\Pi(p)$ has vanishing time 
component in the particle rest frame, as it is apparent from (\ref{luba}). Before 
working out eq.~(\ref{pola}), we should introduce some important notions about the 
construction of physical states. We will stick to the notation of ref.~\cite{moussa}.

As it is well known, the Pauli-Lubanski vector fulfills the commutation relations:
\begin{eqnarray}
&& [\widehat W_\mu, \widehat P_\nu] = 0 \\ \nonumber
&& [\widehat W_\mu, \widehat W_\nu] = -i \sum_{\rho \sigma} 
 \epsilon_{\mu \nu \rho \sigma} \widehat W^\rho \widehat P^\sigma \\ \nonumber
&& \widehat W \cdot \widehat P = 0
\end{eqnarray}
Hence, if the ket $\ket{p}$ is an eigenvector of $\widehat P$, so is $\widehat W \ket{p}$. 
The restriction of $\widehat W$ to the eigenspace labelled by four-momentum $P$ is defined 
as $\widehat W(p)$. Since $\widehat W(p) \cdot p = 0$, this four-vector operator can be 
decomposed onto three orthonormal spacelike four-vectors $n_1(p),n_2(p),n_3(p)$ 
which form a basis of the Minkowski space with the unit vector 
$\hat p = p/\sqrt{p^2}$:
\begin{equation}\label{luba2}
  \widehat W(p) = \sum_{i=1}^3 \widehat W_i(p) n_i(p)
\end{equation}
It can be shown that the operators:
\begin{equation}\label{spins}
  \widehat S_i(p) = \widehat W_i(p)/m
\end{equation}
form an SU(2) algebra and are the actual relativistic generalization of the spin
angular momentum. The third component $\widehat S_3(p)$ can be diagonalized along with  
$\widehat S^2 = -\widehat W^2/m^2$ which is a Casimir of the full group \poinc, with 
corresponding eigenvalues $\sigma$ and $S(S+1)$. With a suitable choice of $n_i(p)$, 
i.e.:
\begin{equation}\label{choi}
  n_i (p) = [p]e_i  \qquad [p] \equiv {\sf R}_3(\varphi) {\sf R}_2 (\theta) 
 {\sf L}_3(\xi)
\end{equation}
$e_i$ being the unit vectors of spacial axes and $[p]$ being a Lorentz transformation 
bringing the timelike vector $p_0 = (m,0,0,0)$ into the four-momentum $p$ with 
polar coordinates $\xi,\theta,\varphi$; the eigenvalue $\lambda$ has the physical 
meaning of the component of intrinsic angular momentum in the rest frame along 
the direction of particle momentum ${\bf p}$. Thus, with the choice (\ref{choi}), 
$\lambda$ is the helicity in the rest frame and $J$ is, by definition, {\em the} 
spin of the particle. Since, from eqs.~(\ref{spins}), (\ref{luba2}) and (\ref{luba}) 
$\widehat S_i(p_0) = \widehat J_i$, the spin operators in the rest frame coincide with 
the generators of the rotation groups. Finally, single particle states will be 
written as $\ket{p,\sigma}$ with:
\begin{equation} \label{quantiz}
 \widehat{P}\ket{p,\sigma} = p \ket{p,\sigma} \qquad {\rm and} \qquad 
 \widehat{S}_3(p)\ket{p,\sigma} = \sigma \ket{p,\sigma} 
\end{equation}
and normalization: 
\begin{equation}\label{norma}
 \braket{p,\sigma}{q,\tau} = \delta^3({\bf p}-{\bf q})\delta_{\sigma \tau}\;.
\end{equation}
while the transformation of a state $\ket{p,\sigma}$ under a general Lorentz transformation 
${\sf \Lambda}$ reads:
\begin{equation} \label{lorentztr}
\widehat{{\sf \Lambda}}\ket{p,\sigma} = \sum_\tau \ket{{\sf \Lambda}p,\tau} 
D^S_{\tau \sigma}([{\sf \Lambda}p]^{-1}{\sf \Lambda} [p]) 
\sqrt{\frac{(\Lambda p)^0}{p^0}}\; .
\end{equation}

We are now in a position to develop eq.~(\ref{pola}). By using (\ref{luba2}) and 
(\ref{spins}) we get:
\begin{equation}\label{pola2}
 \Pi(p) = \frac{1}{m} \sum_{\sigma,\sigma'} \bra{p,\sigma} \widehat W \ket{p,\sigma'}
  \bra{p,\sigma'} \widehat \rho_\omega(p) \ket{p,\sigma} = \sum_{i=1}^3
  \sum_{\sigma,\sigma'} \bra{p,\sigma} \widehat S_i(p) \ket{p,\sigma'} 
  \bra{p,\sigma'} \widehat \rho_\omega(p) \ket{p,\sigma} n_i(p)
\end{equation}
Two different matrices appear in the above equation. By definition:
\begin{equation}\label{spinmatrix}
 \bra{p,\sigma} \widehat S_i(p) \ket{p,\sigma'} = D^S_{\sigma\sigma'}(J_i)
\end{equation}
are the matrices of the SU(2) generators $J_i$ in the representation labeled by the
particle spin $S$. Furthermore, according to (\ref{rhop}):
\begin{equation}\label{spindens}
 \bra{p,\sigma'} \widehat \rho_\omega(p) \ket{p,\sigma} = 
 \frac{1}{\frac{1}{(2\pi)^3} \int_V \d^3 \x \; 
  \e^{\omegav \cdot({\bf x} \times {\bf p})/T} \,\tr D^{S}({\sf R}_{\hat{\omegav}}
 (i\omega/T))} \bra{p,\sigma'} \exp[\omegav \cdot \widehat{\bf j}/T]\Pro_V \ket{p,\sigma} 
\end{equation}
As to the rightmost matrix element, we will restore the vector $\phiv=i\omegav/T$
and calculate it for real $\phiv$ (or imaginary $\omegav$), then making an analytic 
continuation to imaginary $\phiv$, the same way we did at the end of 
Sect.~\ref{rotating}. Therefore:
\begin{equation}
 \bra{p,\sigma'} \exp[\omegav \cdot \widehat{\bf j}/T]\Pro_V \ket{p,\sigma}
 = \bra{p,\sigma'} \exp[-i \phiv \cdot \widehat{\bf j}]\Pro_V \ket{p,\sigma}
 = \bra{p,\sigma'} {\widehat {\sf R}}_{\hat \phiv}(\phi) \Pro_V \ket{p,\sigma}
\end{equation}
The latter matrix element has been calculated in ref.~\cite{micro2}
\footnote{It can be easily obtained from eq.~(55) in the reference}:
\begin{eqnarray}\label{matrix2}
&& \bra{p,\sigma'}{\widehat {\sf R}}_{\hat \phiv}(\phi) \Pro_V \ket{p,\sigma} = 
 \int \d^3\p' \; \delta^3({\sf R}_{\hat \phiv}(\phi){\bf p}'-{\bf p}) 
  \frac{1}{(2\pi)^3} \int_V \d^3 \x \; \e^{i {\bf x}\cdot({\bf p}-{\bf p}')} 
  \nonumber \\
 && \qquad \qquad \qquad \qquad \qquad 
  \times \frac{1}{2} \left(D^S([p]^{-1}{\sf R}_{\hat \phiv}(\phi)[p])
   +D^S([p]^{\dagger}{\sf R}_{\hat \phiv}(\phi)[p]^{\dagger-1})\right)_{\sigma'\sigma}
 \bra{0} \Pro_V \ket{0} \nonumber \\
&& =\frac{1}{(2\pi)^3} \int_V \d^3 \x \; \e^{i {\bf x}\cdot({\bf p}-
 {\sf R}_{\hat \phiv}(\phi)^{-1}{\bf p})} 
  \frac{1}{2} \left(D^S([p]^{-1}{\sf R}_{\hat \phiv}(\phi)[p])
  +D^S([p]^{\dagger}{\sf R}_{\hat \phiv}(\phi)[p]^{\dagger-1})\right)_{\sigma'\sigma}
 \bra{0} \Pro_V \ket{0} 
\end{eqnarray}
In this equation, $[p]$ is now meant as an element of the universal covering group 
of Lorentz group, i.e. SL(2,C), so that the notation $[p]^\dagger$ becomes meaningful 
as well as its finite-dimensional representation matrix $D^S([p])$ corresponding to
the Lorentz group representation usually labelled as $(S,0)$ (the $(0,S)$ being the
$D^S([p])^{\dagger -1}$). The factor $\bra{0} \Pro_V \ket{0}$ is immaterial and also 
becomes 1 in the large volume limit as $\Pro_V \to {\rm I}$. 

If $\phi \ll 1$, as it was supposed to be at the beginning of this section, then: 
$$
 i {\bf x}\cdot({\bf p}-{\sf R}_{\hat \phiv}(\phi)^{-1}{\bf p}) \simeq
 - i \phiv \cdot({\bf x} \times {\bf p})
$$
and we can rewrite eq.~(\ref{matrix2}) as:
\begin{equation}\label{matrix3}
 \bra{p,\sigma'}{\widehat {\sf R}}_{\hat \phiv}(\phi) \Pro_V \ket{p,\sigma} \simeq
 \frac{1}{(2\pi)^3} \int_V \d^3 \x \; \e^{-i \phiv \cdot({\bf x} \times {\bf p})} 
  \frac{1}{2} \left(D^S([p]^{-1}{\sf R}_{\hat \phiv}(\phi)[p])
  +D^S([p]^{\dagger}{\sf R}_{\hat \phiv}(\phi)[p]^{\dagger-1})\right)_{\sigma'\sigma}
\end{equation}
We can now make an analytical continuation to imaginary $\phiv$ or real $\omegav$
and obtain, by using (\ref{spindens}) and  (\ref{matrix3}):
\begin{equation}\label{spindens2}
 \bra{p,\sigma'} \widehat \rho_\omega(p) \ket{p,\sigma} =
  \frac{1}{2 \tr D^S({\sf R}_{\hat\omegav}(i\omega/T))} 
 \left(D^S([p]^{-1}{\sf R}_{\hat \omegav}(i\omega/T)[p])
  +D^S([p]^{\dagger}{\sf R}_{\hat\omegav}(i\omega/T)[p]^{\dagger-1})\right)_{\sigma'\sigma}   
\end{equation}
This matrix is hermitian and has trace equal to 1, as required for a good density 
matrix. Hermiticity can be shown by taking advantage of a remarkable feature of
SL(2,C) representation \cite{moussa}:
$$
 D^S (A^\dagger) = D^S (A)^\dagger
$$
and taking into account that $D^S({\sf R}_{\hat \omegav}(i\omega/T))$ is hermitian.

Plugging (\ref{spindens2}) into (\ref{pola2}), using (\ref{spinmatrix}) and carrying 
out the momentum integration, the final expression of the polarization four-vector 
for a particle of momentum $p$ in the observer frame is finally achieved:
\begin{eqnarray}\label{polafin}
 \Pi(p)^\mu &=&  \frac{1}{2 \tr D^S({\sf R}_{\hat\omegav}(i\omega/T))} 
  \sum_{i=1}^3 \tr \left[ D^S(J_i [p]^{-1} {\sf R}_{\hat \omegav}(i\omega/T)[p])
  +D^S(J_i[p]^{\dagger}{\sf R}_{\hat\omegav}(i\omega/T)[p]^{\dagger-1})\right] n_i(p) 
  \nonumber \\
 &=&  \frac{1}{2 \tr D^S({\sf R}_{\hat\omegav}(i\omega/T))}\sum_{i=1}^3 
  \tr \left[ D^S(J_i [p]^{-1} {\sf R}_{\hat \omegav}(i\omega/T)[p])
  +D^S(J_i[p]^{\dagger}{\sf R}_{\hat\omegav}(i\omega/T)[p]^{\dagger-1})\right]
  ([p]e_i)^\mu
\end{eqnarray}
The proper polarization four-vector, in the particle rest-frame, has components
$([p]^{-1} \Pi(p))^\mu$, that is:
\begin{equation}\label{properpolafin}
 \Pi_0(p)^\mu =
 \frac{1}{2 \tr D^S({\sf R}_{\hat\omegav}(i\omega/T))}  \sum_{i=1}^3
  \tr \left[ D^S(J_i [p]^{-1}{\sf R}_{\hat \omegav}(i\omega/T)[p])
  +D^S(J_i[p]^{\dagger}{\sf R}_{\hat\omegav}(i\omega/T)[p]^{\dagger-1})\right] \delta^\mu_i
\end{equation}
which has vanishing time component, as required.

The equations (\ref{polafin}) and (\ref{properpolafin}) are the general analytical 
expression of the polarization for a particle with spin $S$. We will now develop them
for the most interesting cases of spin 1/2 and spin 1. Before doing this, we first 
observe that the Lorentz transformation $[p]$ to calculate the polarization can be 
chosen arbitrarily. In fact, if $[p]' \ne [p]$ also transforms the timelike vector 
$p_0 = (m,0,0,0)$ into $p$, then $[p]^{-1}[p]'$ is a pure rotation ${\sf R}$ as it 
leaves $p_0$ invariant. Hence, the polarization four-vector $\Pi(p)$ defined with 
$[p]'$ becomes:
\begin{eqnarray}\label{polaalt}
 \Pi(p) &\propto& \sum_{i=1}^3 \frac{1}{2} \tr \left[ D^S(J_i [p]'^{-1}
  {\sf R}_{\hat \omegav}(i\omega/T)[p]')
  +D^S(J_i[p]'^{\dagger}{\sf R}_{\hat\omegav}(i\omega/T)[p]'^{\dagger-1})\right]
  ([p]'e_i) \nonumber \\
 &=& \sum_{i=1}^3 \frac{1}{2} \tr \left[ D^S(J_i {\sf R}^{-1} [p]^{-1}
  {\sf R}_{\hat \omegav}(i\omega/T)[p] {\sf R})
  +D^S(J_i {\sf R}^{-1} [p]^{\dagger}{\sf R}_{\hat\omegav}(i\omega/T) [p]^{\dagger-1}
   {\sf R})\right]([p]{\sf R}e_i) \nonumber \\
 &=& \sum_{i=1}^3 \frac{1}{2} \tr \left[ D^S({\sf R} J_i {\sf R}^{-1} [p]^{-1}
  {\sf R}_{\hat \omegav}(i\omega/T)[p] )
  +D^S({\sf R} J_i {\sf R}^{-1} [p]^{\dagger}{\sf R}_{\hat\omegav}(i\omega/T) [p]^{\dagger-1}
   )\right]([p]{\sf R}e_i) \nonumber \\ 
 &=& \sum_{i=1}^3 \frac{1}{2} \tr \left[ D^S(J_{{\sf R}e_i} [p]^{-1}
  {\sf R}_{\hat \omegav}(i\omega/T)[p] )
  +D^S(J_{{\sf R}e_i} [p]^{\dagger}{\sf R}_{\hat\omegav}(i\omega/T) [p]^{\dagger-1}
   )\right]([p]{\sf R}e_i) \nonumber \\
\end{eqnarray}
where we used the unitarity of ${\sf R}$. The last expression in (\ref{polaalt}) 
is apparently equal to the one in (\ref{polafin}).

\subsection{Spin S=1/2}

Choosing for $[p]$ a pure Lorentz boost, we have \cite{moussa}:
\begin{equation}\label{ferm1}
  D^{1/2}([p]) = D^{1/2}([p])^\dagger = 
 \frac{m + \varepsilon + {\sigmav}\cdot{\bf p}}{\sqrt{2m(m+\varepsilon)}}
\end{equation}
and
\begin{equation}\label{ferm2}
  D^{1/2}([p])^{-1} = D^{1/2}([p])^{\dagger-1} = 
 \frac{m + \varepsilon - {\sigmav}\cdot{\bf p}}{\sqrt{2m(m+\varepsilon)}}
\end{equation}
where $\sigmav$ are the Pauli matrices. Therefore:
\begin{equation}\label{trace1}
 \tr \left[ D^{1/2}(J_i [p]^{-1}{\sf R}_{\hat \omegav}(i\omega/T)[p])\right] = 
 \tr \left[ \frac{\sigma_i}{2} 
 \frac{m + \varepsilon - {\sigmav}\cdot{\bf p}}{\sqrt{2m(m+\varepsilon)}}
 {\sf R}_{\hat \omegav}(i\omega/T)\frac{m + \varepsilon + {\sigmav}\cdot{\bf p}}
 {\sqrt{2m(m+\varepsilon)}})\right] 
\end{equation}
whereas, because of (\ref{ferm1}) and (\ref{ferm2}):
\begin{equation}\label{trace2}
 \tr \left[ D^{1/2}(J_i [p]^\dagger {\sf R}_{\hat \omegav}(i\omega/T)
 [p]^{\dagger-1}) \right] = \tr \left[ \frac{\sigma_i}{2} 
 \frac{m + \varepsilon + {\sigmav}\cdot{\bf p}}{\sqrt{2m(m+\varepsilon)}}
 {\sf R}_{\hat \omegav}(i\omega/T)\frac{m + \varepsilon - {\sigmav}\cdot{\bf p}}
 {\sqrt{2m(m+\varepsilon)}})\right]
\end{equation}
i.e. it is obtained from (\ref{trace1}) by reflecting ${\bf p}$.
Since:
\begin{equation}
 D^{1/2}({\sf R}_{\hat \omegav}(i\omega/T)) = {\rm I} \cos \frac{i\omega}{2T} 
 - i \sigmav \cdot \hat\omegav \sin \frac{i\omega}{2T}
 = {\rm I} \cosh \frac{\omega}{2T} + \sigmav \cdot \hat\omegav 
  \sinh \frac{\omega}{2T}
\end{equation}
we have, for (\ref{trace1})
\begin{eqnarray}\label{interm1}
&& \frac{m + \varepsilon - {\sigmav}\cdot{\bf p}}{\sqrt{2m(m+\varepsilon)}}
 \left( {\rm I} \cosh \frac{\omega}{2T} + \sigmav \cdot \hat\omegav \sinh 
 \frac{\omega}{2T} \right)
 \frac{m + \varepsilon + {\sigmav}\cdot{\bf p}}{\sqrt{2m(m+\varepsilon)}} \nonumber \\
 && = {\rm I} \cosh \frac{\omega}{2T} + \sinh \frac{\omega}{2T} \left[
 \frac{m+\varepsilon}{2m} \sigmav \cdot \hat \omegav  + \frac{i}{m} \sigmav \cdot 
 (\hat\omegav \times {\bf p}) - \frac{\hat \omegav \cdot {\bf p} \sigmav \cdot {\bf p} - 
 \sigmav \cdot ({\bf p} \times (\hat \omegav \times {\bf p} ))}{2m(\varepsilon+m)} 
 \right]
\end{eqnarray}
Now we have to take the sum of the two traces (\ref{trace1}) and (\ref{trace2}) 
implying that all terms which change sign in a reflection of the momentum ${\bf p}$
in (\ref{interm1}) vanish. Therefore, we are left with:
\begin{eqnarray}\label{tedious}
&& 2 \tr \left\{ \frac{\sigma_i}{2} \, \left[{\rm I} \cosh \frac{\omega}{2T} + 
 \sinh \frac{\omega}{2T} \left[
 \frac{m+\varepsilon}{2m} \sigmav \cdot \hat \omegav - 
 \frac{\hat \omegav \cdot {\bf p} \sigmav \cdot {\bf p} - 
 \sigmav \cdot ({\bf p} \times (\hat \omegav \times {\bf p} ))}{2m(\varepsilon+m)} \right]
 \right] \right\} \nonumber \\
&&= \sinh \frac{\omega}{2T} \left[ \frac{m+\varepsilon}{m} \hat \omegav_i 
- \frac{\hat \omegav \cdot {\bf p} {\bf p}_i - 
 ({\bf p} \times (\hat \omegav \times {\bf p}))_i}{m(\varepsilon+m)} \right]
\end{eqnarray}
Taking into account that:
\begin{equation}\label{trace12}
 \tr D^{1/2}({\sf R}_{\hat \omegav}(i\omega/T)) = 2 \cosh \frac{\omega}{2T} 
\end{equation}
we can obtain the polarization vector in the rest-frame putting (\ref{tedious}) and
(\ref{trace12}) into eq.~(\ref{properpolafin}):
\begin{eqnarray}\label{proper}
 \Piv_0 &=& \frac{1}{4} \tanh \frac{\omega}{2T} 
 \left[ \frac{m+\varepsilon}{m} \hat \omegav - \frac{\hat \omegav \cdot {\bf p} 
 {\bf p} - {\bf p} \times (\hat \omegav \times {\bf p})}{m(\varepsilon+m)} 
 \right] \nonumber \\
 &=& \frac{1}{4} \tanh \frac{\omega}{2T} 
 \left[ \frac{m+\varepsilon}{m} \hat \omegav - \frac{\hat \omegav \cdot {\bf p} 
 {\bf p} - \p^2 \hat \omegav + \hat \omegav \cdot {\bf p} {\bf p}}{m(\varepsilon+m)} 
 \right] \nonumber \\
 &=& \frac{1}{2} \tanh \frac{\omega}{2T} 
 \left[ \frac{\varepsilon}{m} \hat \omegav - \frac{\hat \omegav \cdot {\bf p} 
 {\bf p}}{m(\varepsilon+m)} \right] 
\end{eqnarray}
Therefore, unlike in the non-relativistic case, the polarization vector has a 
component along particle momentum in the observer frame. This effect is owing to the
vector nature of the polarization; indeed, the components of polarization in the
observer frame (\ref{polafin}) can be obtained with a general Lorentz boost 
\cite{jackson}:
\begin{equation}\label{boost}
 {\Piv} = {\Piv_0} + \frac{\gamma^2}{\gamma+1} {\betav \cdot \Piv_0} \betav 
 \qquad \qquad \Pi^0 = \gamma \betav \cdot \Piv_0
\end{equation}
with $\betav = {\bf p}/\varepsilon$ and $\gamma = \varepsilon/m$. Therefore, by 
using (\ref{proper}), eq.~(\ref{boost}) becomes:
\begin{equation}\label{boost2}
 {\Piv} = \frac{1}{2} \tanh \frac{\omega}{2T} \frac{\varepsilon}{m} \hat \omegav
 \qquad \qquad \Pi^0 = \frac{1}{2} \tanh \frac{\omega}{2T}\frac{\hat \omegav 
 \cdot {\bf p}}{m}
\end{equation}
Hence, the polarization three-vector is aligned with the angular velocity in the
{\em observer frame} and not in the particle frame. Yet, there is also a time 
component which vanishes in the non-relativistic limit ${\bf p}/m \to 0$, where
the (\ref{boost2}) correctly yields (\ref{nonrela}). The remarkable difference 
with respect to the non-relativistic case (\ref{nonrela}) is that polarization 
now depends on momentum. The longitudinal component, along ${\bf p}$, is:
\begin{equation}
  \Piv_0 \cdot \hat{\bf p} = \frac{1}{2} \tanh \frac{\omega}{2T} 
  \hat{\bf p} \cdot \hat\omegav
\end{equation}
while the component along the rotation axis turns out to be:
\begin{equation}\label{polomega}
  \Piv_0 \cdot \hat{\omegav} = \frac{1}{2} \tanh \frac{\omega}{2T} 
 \left[ \frac{\varepsilon}{m} - \frac{\p^2 (\hat \omegav \cdot \hat {\bf p})^2}
  {m(\varepsilon+m)} \right] 
\end{equation}
which shows a very interesting feature: the polarization is maximal for particles 
with momentum orthogonal to the rotation axis. Furthermore, the polarization increases
with energy, being proportional to $\varepsilon/m$. Of course this behaviour cannot
go on indefinitely because polarization cannot exceed $1/2$ in any direction; since 
$\omega/T \ll 1$, it is seen from eq.~(\ref{polomega}) that something must happen 
when $\varepsilon \sim 2 m T /\omega$. Indeed, we have pointed out in Sect.~\ref{spect}
that when momentum is of the order of $J/R$, the saddle point expansion of partition 
function at fixed $J$ can no longer be independent of ${\rm p}$, and $\omega$ becomes 
in fact a function of $p$, that is $\omega(p)$, for the spectrum (\ref{genspect}) expansion. 
This has some impact on the polarization vector: at some large momentum the dependence
of $\omega$ on $p$ should restore the natural $1/2$ bound.

\subsection{Spin S=1}

The calculation of the polarization for massive spin 1 particles is more involved
than for spin $1/2$ and we have carried it out by performing explicitely the 
multiplication of matrices of  SL(2,C) $D^1$ representation. Since eventually one 
has to calculate traces, the choice of the basis for the matrices is arbitrary and 
we have written $[p]={\sf R}_3(\varphi){\sf R}_2(\theta){\sf L}_3(\xi)$, being 
$\cosh \xi=\varepsilon/m$, in the cartesian basis for the $D^1$ representation space: 
\begin{eqnarray}\label{spin1mat}
&& D^1({\sf R}_3(\varphi))
 = \left( \begin{array}{ccc} \cos \varphi  & -\sin \varphi & 0 \\
                             \sin \varphi  & \cos \varphi  & 0 \\
                              0            &  0             & 1 \\
         \end{array} \right) 
 \qquad \qquad
 D^1({\sf R}_2(\theta))
 = \left( \begin{array}{ccc} \cos \theta  & 0  &  \sin \theta  \\
                              0           & 1  &       0       \\
                             -\sin \theta  & 0  &  \cos \theta  \\
         \end{array} \right) 
\nonumber \\
&& D^1({\sf L}_3(\xi))= \exp[-i \xi K_3] = \exp[ \xi J_3] = D^1({\sf R}_3(i\xi))
 = \left( \begin{array}{ccc} \cosh \xi    & -i\sinh \xi    & 0 \\
                             i \sinh \xi  & \cosh \xi      & 0 \\
                              0           &  0            & 1 \\
         \end{array} \right) 
\nonumber \\
&& D^1({\sf R}_{\hat \omegav}(i \omega/T))= \exp[\hat\omegav/T \cdot D^1({\bf J})]
\qquad \qquad  D^1(J_i)_{jk} = -i \epsilon_{ijk} 
\end{eqnarray}
Multiplying the matrices in eq.~(\ref{spin1mat}) and calculating the traces, according 
to eq.~(\ref{properpolafin}), one gets:
\begin{equation}
 \Piv_0 = \frac{2 \sinh (\omega/T)}{2 \cosh (\omega/T) + 1} 
  \left[ \frac{\varepsilon}{m} \hat \omegav - \frac{\hat \omegav \cdot {\bf p} 
 {\bf p}}{m(\varepsilon+m)} 
 \right]
\end{equation}
where the denominator $2 \cosh (\omega/T) + 1$ is the trace of the matrix 
$D^1({\sf R}_{\hat \omegav}(i \omega/T))$.
Therefore, the kinematical structure of the polarization vector is the same as in 
the spin 1/2 case, which is a reasonable outcome; in fact, the properties of Lorentz
transformation of a polarization vector should not depend on the particle spin 
itself.

In the case of spin 1 it is also interesting to calculate the fraction of the 
$00$ component of the density matrix (\ref{spindens2}). The calculation can be done
quickly by noting that:
\begin{equation}\label{align1}
  \rho_{\omega\,00}(p) = \tr {P_3 \widehat \rho} = 
  \sum_\sigma \bra{p,\sigma} P_3 \widehat \rho_\omega(p) \ket {p,\sigma}
\end{equation}
$P_3$ being the projector onto the state $\ket{p,0}$. Written in the cartesian basis
the matrix corresponding to $P_3$ is simply:
\begin{equation}
 P_3 = \left( \begin{array}{ccc} 0  &  0  &  0 \\
                                 0  &  0  &  0 \\
                                 0  &  0  &  1 \\
         \end{array} \right) 
\end{equation}
because of the choice of $z$ axis as quantization axis, according to (\ref{quantiz})
Therefore, by plugging eq.~(\ref{spindens2}) into eq.~(\ref{align1}):
\begin{equation}
  \rho_{\omega\,00}(p) = 
   \frac{1}{2 \tr D^1({\sf R}_{\hat\omegav}(i\omega/T))} 
   \tr \left[ P_3 D^1([p]^{-1}{\sf R}_{\hat \omegav}(i\omega/T)[p])
   + P_3 D^1([p]^{\dagger}{\sf R}_{\hat\omegav}(i\omega/T)[p]^{\dagger-1})\right]
\end{equation}
yielding:
\begin{equation}
  \rho_{\omega\,00}(p) = \frac{1}{2 \cosh (\omega/T) + 1}
 \left[ \cosh(\omega/T) + \frac{({\bf p} \cdot \omegav)^2}{\p^2 \omega^2} 
 \left( 1 - \cosh(\omega/T) \right) \right]
\end{equation}
For small $\omega/T$, we have:
\begin{equation}
  \rho_{\omega\,00}(p) \simeq \frac{1}{3} + 
  \frac{1-3({\hat {\bf p}}\cdot{\hat \omegav})^2}{18} \frac{\omega^2}{T^2}
\end{equation}
%

\subsection{Discussion}

In both spin 1/2 and spin 1 cases, we have seen that the proper polarization vector
has the same structure, so that we can easily generalize the above results to any 
spin:
\begin{equation}\label{anyspin}
 \Piv_0 = \frac{\sum_{n=-S}^S n \e^{n\omega/T}}{\sum_{n=-S}^S \e^{n\omega/T}} 
  \left[ \frac{\varepsilon}{m} \hat \omegav - \frac{\hat \omegav \cdot {\bf p} 
 {\bf p}}{m(\varepsilon+m)} 
 \right]
\end{equation}
This non-vanishing polarization means that spin states are not evenly populated 
in an equilibrated thermodynamical system with finite, macroscopic, angular momentum.
Of course reaching equilibrium for the spin degrees of freedom implies that a
small interaction should exists involving particle spin, e.g. a spin-orbit coupling.
Yet, for the ideal relativistic gas, this interaction is assumed to be negligible
in comparison with the pure kinematical effect of angular momentum conservation. 
This can be rephrased in the rotating frame as follows: the actual 
interaction involving spin is negligible with respect to the spin-rotation coupling. 
Since a macroscopic thermodynamical system can be conceptually divided into elementary 
fluid cells, and each cell participating in the rigid rotation is an accelerated system, 
for our result to be consistent with locality, one has to conclude that, in general, 
acceleration involves a polarization expressed by (\ref{anyspin}) where $\omegav$ is 
to be interpreted as a local vector field involving local quantities such as velocity 
and acceleration of the cell. 

\section*{Acknowledgments}

We are grateful for interesting discussions to R. Jaffe, G. Longhi, L. Lusanna, 
K. Rajagopal, H. Satz, D. Seminara. We thank Galileo Galilei Institute for hospitality.

\appendix

\section*{APPENDIX A - Entropy for a spinning system}

The entropy of a system with finite angular momentum ${\bf J}$ can be calculated 
easily from the expression of the probability of a state with energy $E$, 
vanishing momentum, charge $Q$ and fixed ${\bf J}$ in the grand-canical ensemble:
\begin{equation}
  p = \frac{1}{Z_J} \exp[-E/T + \mu Q/T]
\end{equation}
so that, by using (\ref{legendre})
\begin{equation}\label{entropy}
 S = - \underset{states with fixed {\bf J}}{\sum} p \log p  
   = \frac{\langle E \rangle}{T} - \frac{\mu \langle Q \rangle}{T} +¨ \log Z_J = 
     \frac{U}{T} - \frac{\mu \langle Q \rangle}{T} -
     \frac{\omegav \cdot {\bf J}}{T} + \log Z_\omega
\end{equation}
which is the known expression of entropy for a rotating system. The logarithm
of the partition function can then be identified with the integral of the
pressure:
\begin{equation}
 \log Z_\omega = \frac{1}{T} \int_V \d^3 \x \; p({\bf x})
\end{equation}
where, unlike in familiar cases, the pressure is not uniform due to rotation. From 
the entropy expression:
\begin{equation}
 TS = U - {\mu \langle Q \rangle} - \omegav \cdot {\bf J} + 
 \int_V \d^3 \x \; p({\bf x})
\end{equation}
we can derive the relation \cite{landau}
\begin{equation}\label{jderiv}
 \frac{\partial S}{\partial {\bf J}}\Big|_{V,U,Q} = -\frac{\omegav}{T}
\end{equation}
%

\section*{APPENDIX B - Equilibrium configuration of a relativistic rotating 
system}

We will now show that a relativistic macroscopic system with finite angular momentum 
at equilibrium must be rigidly rotating by generalizing an argument by Landau \cite{landau}
for non-relativistic system.

Let us consider a generic isolated hdyrodynamical system with velocities ${\bf v}_i$
$i$ being the label of hydrodynamical cells. In order to find the equilibrium
configuration the entropy should be maximized with the constraint of energy,
momentum and angular momentum conservation. Therefore, we have to find the extremum
points of:
\begin{equation}\label{entropy2}
 \sum_i S_i - \beta (\sum_i E_i - E_0) + \betav \cdot {\sum_i {\bf P}_i} + 
 \beta \omegav \cdot (\sum_i {\bf x_i} \times {\bf P}_i - {\bf J})  
\end{equation}
where we have taken the total momentum vanishing, i.e. we are working in the system's
rest frame; $E_0$ is the total energy and ${\bf J}$ the total angular momentum;
$\beta$, $\betav$ and $\omegav/T$ are Lagrange multipliers enforcing the conservation
laws. Since the entropy is a relativistic invariant, it can only depend on the mass 
of the cell, i.e. $S_i = S_i (\sqrt{E^2_i - {\bf P}_i^2})$. In order to find the 
equilibrium configuration, one has to maximize (\ref{entropy2}) with respect to
each ${\bf P}_i$ and $E_i$. Thus:
\begin{equation}\label{enderiv}
 \frac{\partial S_i}{\partial E_i} = \frac{E_i}{M_i} \frac{\partial S_i}{\partial M_i}
 = \beta \qquad \forall i  
\end{equation}
and:
\begin{equation}\label{pderiv}
 -\frac{\partial S_i}{\partial {\bf P_i}} = \frac{{\bf P}_i}{M_i} 
 \frac{\partial S_i}{\partial M_i} = \betav + \beta \omegav \times {\bf x}_i 
 \qquad \forall i  
\end{equation}
Taking into account that $\partial S_i/\partial M_i$ is the inverse of the 
proper temperature $T_i$ of the $i$th cell by definition, eq.~(\ref{enderiv}) 
implies that:
\begin{equation}\label{temperature}
  \frac{\gamma_i}{T_i} = {\rm constant} = \beta \equiv \frac{1}{T}
\end{equation}
where $T$ is defined the {\em global} temperature of the system. Plugging 
(\ref{temperature}) into (\ref{pderiv}) we get:
\begin{equation}\label{velocit1}
  \frac{\gamma_i {\bf v}_i}{T_i} = \frac{{\bf v}_i}{T} = \betav + 
  \frac{\omegav \times {\bf x}_i}{T}
\end{equation}
For the total momentum to vanish, the vector $\betav$ should be 0 as well and we
are left with:
\begin{equation}\label{velocit2}
  {\bf v}_i = \omegav \times {\bf x}_i
\end{equation}
that is a rigid rotation around the axis $\omegav$. To show that $T = 1/\beta$ is 
in fact the global temperature of the sytem, one has to consider that at equilibrium,
using formula (\ref{entropy}) in Appendix A for the entropy of the cell $i$:
\begin{equation}
 S \equiv \sum_i S_i = \sum_i \frac{M_i}{T_i} + \ldots = \sum_i \frac{\gamma_i M_i}{T}
 + \ldots =  \sum_i \frac{E_i}{T} = \frac{M}{T} + \ldots
\end{equation}
$M$ being the mass of the system, because ${\bf P} = 0$. Therefore:
$$
 \frac{\partial S}{\partial M} = \frac{1}{T}
$$
showing that $T$ is the actual temperature of the system. The relation (\ref{temperature})
can be rewritten as, by using (\ref{velocit2}):
\begin{equation}
  T = T_i \sqrt {1 - \| \omegav \times {\bf x}_i \|^2}
\end{equation}
in accordance with eq.~(\ref{temp}). The involved physical meaning is as follows: 
the temperature measured by a thermometer at rest in the observer frame, is {\em lower} 
than that measured by a comoving thermometer, i.e. at rest in the cell frame. This 
is a pure relativistic effect which has no correspondance in classical thermodynamics. 

The above argument can be extended to include the intrinsic angular momentum 
contribution ${\bf J}_i$ of each cell to ${\bf J}$. Eq.~(\ref{entropy2}) now reads:  
\begin{equation}\label{entropy3}
 \sum_i S_i (\sqrt{E_i^2 - {\bf P}_i^2},{\bf J}_i) - 
 \beta (\sum_i E_i - E_0) + \betav \cdot {\sum_i {\bf P}_i} +
 \beta \omegav \cdot (\sum_i {\bf x_i} \times {\bf P}_i + {\bf J}_i - {\bf J})  
\end{equation}
The conclusions are similar, with the additional condition:
$$
 \frac{\partial S_i}{\partial {\bf J_i}} = -\frac{\omegav}{T}
$$
in accordance with eq.~(\ref{jderiv}).



\end{document}